\pdfoutput=1
\documentclass[prd,aps,preprintnumbers,showpacs,10pt,twocolumn,nofootinbib]{revtex4}
\usepackage{amsfonts,amsmath,amssymb,bm,eucal,eufrak,slashed}
\usepackage[english]{babel}
\usepackage[latin1]{inputenc}
\usepackage[T1]{fontenc}
\usepackage{ae,aecompl}
\usepackage{graphicx,graphics,epsfig,epstopdf,subfigure,color,psfrag}
\usepackage{hyperref}

\begin{document}
\preprint{BARI-TH/629-2010}
\preprint{DCPT/10/138}
\preprint{IPPP/10/69}
\title{Holography, Heavy-Quark Free Energy, and the QCD Phase Diagram}
\author{Pietro Colangelo}  
\affiliation{Istituto Nazionale di Fisica Nucleare, Sezione di Bari, Italy}
\author{Floriana Giannuzzi}
\affiliation{Dipartimento di Fisica, Universit\`a degli Studi di Bari\\ 
and Istituto Nazionale di Fisica Nucleare, Sezione di Bari, Italy}
\author{Stefano Nicotri}
\affiliation{Institute for Particle Physics Phenomenology, Department of Physics, Durham University, Durham DH1 3LE, United Kingdom}
\begin{abstract}
We  use gauge/string duality to investigate the free energy of two static color sources (a heavy quark-antiquark pair) in a Yang-Mills theory in strongly interacting  matter, varying temperature and chemical potential.
The  dual space geometry is Anti-de Sitter with a charged black-hole  to describe finite  temperature and density in the boundary theory, and we also include a background 
warp factor to generate confinement.
The resulting deconfinement line in the $\mu-T$ plane is similar to the one obtained by lattice  and effective models of QCD.
\end{abstract}
\maketitle

Consider QCD in a four dimensional Euclidean spacetime and in nuclear matter, and two static color sources, an infinitely heavy quark and an infinitely heavy antiquark, at distance $r$ from each other. 
It is interesting to investigate how the free energy of such a system behaves against variations of temperature and chemical potential.

The study of a strongly coupled Yang-Mills theory, such as QCD, is a challenge in spite of the methods developed so far to deal with it (lattice simulations, models and effective field theories).
The formulation of the gauge/gravity (or Anti-de Sitter/conformal field theory) correspondence \cite{Maldacena:1997re,Witten:1998qj,Gubser:1998,Witten:1998zw} has suggested to face this problem through the identification of a suitable higher dimensional gravity dual.  
Due to  the strong/weak nature of the duality, the gravity dual is a weakly coupled theory defined in a higher dimensional curved spacetime; since QCD is not conformal, a mechanism for breaking conformal invariance must  be included in the dual model. 

This holographic framework can be adopted to analyze finite temperature and density effects. 
In a bottom-up approach, we use the soft wall $AdS$/QCD model, a five dimensional model formulated on  $AdS_5$ spacetime, in which linear confinement at zero temperature and density is obtained by inserting a background  
warp factor
\cite{Andreev:2006vy,Andreev:2006ct,Karch:2006pv},
bringing
a mass scale related to $\Lambda_{\mbox{\tiny QCD}}$.
To describe the boundary theory at finite temperature, a black-hole is included in the five dimensional space, whose horizon position represents  the (inverse) temperature \cite{Witten:1998zw}.

On the other hand, in QCD the effect of finite quark density is introduced by adding the term $J_D=\mu\,\psi^\dagger(x) \psi(x)$ to the Lagrangian in the generating functional, so that the chemical potential $\mu$ appears as the source of the quark density operator. 
According to the $AdS$/CFT  correspondence, the source of a QCD operator in the generating functional is the boundary value of a  dual field in the bulk; therefore, the chemical potential can be considered as the boundary value of the time component of a $U(1)$ gauge field $A_M$ dual to the vector quark current.  
Under the ansatz $A_0 = A_0(z)$ ($z$ is the fifth holographic coordinate) and $A_i = A_z = 0 \,\, (i = 1, 2 , 3)$, one can find a solution of the equations of motion of a $5D$ gravity action with negative cosmological constant interacting with an electromagnetic field: the solution is known as the $AdS$/Reissner-Nordstr\"om black-hole metric, and describes a charged black hole interacting with the electromagnetic field  \cite{johnson,Lee:2009bya}.
We use this (Euclidean) metric in our model:
\begin{eqnarray}\label{metric}
ds^2 & = & \frac{R^2\,e^{c^2 z^2}}{z^2}\left(f(z)dt^2 + d\bar x ^2 + \frac{dz^2}{f(z)}\right)\,\,,\nonumber\\
f(z) & = & 1-\left(\frac{1}{z_h^4}+q^2z_h^2\right)z^4+q^2z^6\,\,,
\end{eqnarray}
where $R$ is the $AdS$ radius, $q$ the black-hole charge and $z_h$ the position of the horizon, defined by the condition $f(z_h)=0$. 
The $e^{c^2z^2}$ term, characterizing the soft wall model, distorts the metric
and brings the mass scale $c$ \cite{Andreev:2006ct,Andreev:2010bv}.
The positive sign in the exponent is chosen, following ref.\cite{Andreev:2006ct}, to obtain a confining $\bar Q Q$ potential at $T=0$. The profile introduced in \cite{Karch:2006pv}, with the negative sign, also produces linear Regge
trajectories and bound states for light hadrons, but does not provides an area law for the Wilson loop, as discussed,  in the framework of light front holography, in \cite{deTeramond:2009xk}
\footnote{Issues related to the choice of the dilaton profile are discussed in \cite{Karch:2010eg}.}.
The boundary $z=0$ represents the four dimensional spacetime on which the Yang-Mills theory is defined. 

The Hawking temperature of the black-hole is 
\begin{equation}
T=\frac{1}{4\pi}\,\left|\frac{df}{dz}\right|_{z=z_h}=\frac{1}{\pi z_h}\,\left(1-\frac{1}{2}\,Q^2\right)\,\,,
\end{equation}
where $Q=q z_h^3$ and $0\leqslant Q\leqslant\sqrt{2}$; moreover, the Euclidean time is a periodic variable, with period $\beta=1/T$ \cite{Witten:1998zw,Herzog:2006ra}. Studies of this kind of models for $q=0$ can be found in \cite{ft}.

The relation between the parameters in the metric \eqref{metric} and the chemical potential can be obtained observing that, on dimensional grounds, the low $z$ behavior of the bulk gauge field $A_0(z)$ is 
\begin{equation}\label{lowz}
A_0(z)=\mu-\eta z^2
\end{equation}
with $\eta=\kappa\,q$ and $\kappa$ a dimensionless parameter in our model.   
Together with the condition that $A_0$ vanishes at  the horizon, $A_0(z_h)=0$, Eq.\eqref{lowz} determines a relation between $\mu$ and the black-hole charge:
\begin{equation}\label{rel}
\mu=\kappa\,q\,z_h^2=\kappa\,\frac{Q}{z_h}\,\,.
\end{equation}
The parameter $c$ in the warp factor in \eqref{metric} sets the scale: in the following, we analyze quantities expressed in units of $c$, and  comment on the numerical results at the end.

\begin{figure}
\subfigure[confined]{
\includegraphics[scale=0.4]{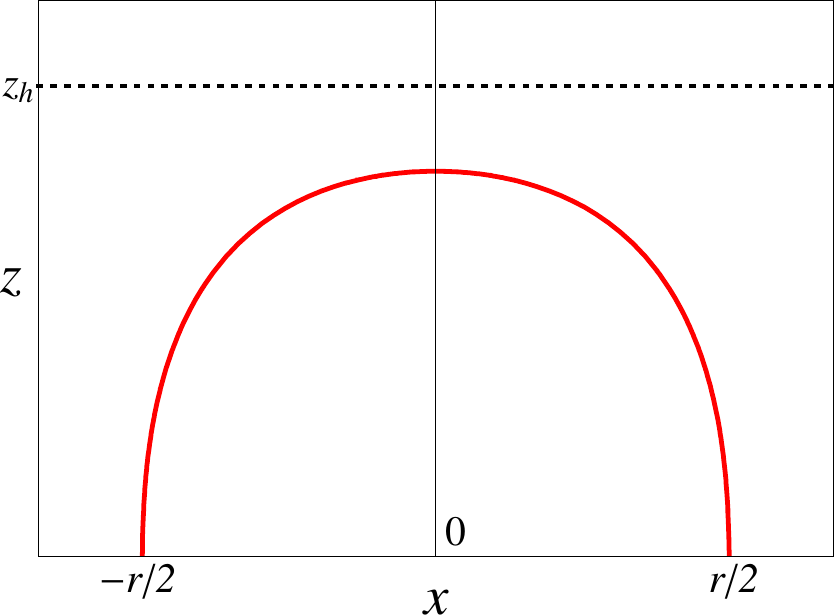}
\label{fig:wsa}
}
\hspace*{.2cm}
\subfigure[deconfined]{
\includegraphics[scale=0.4]{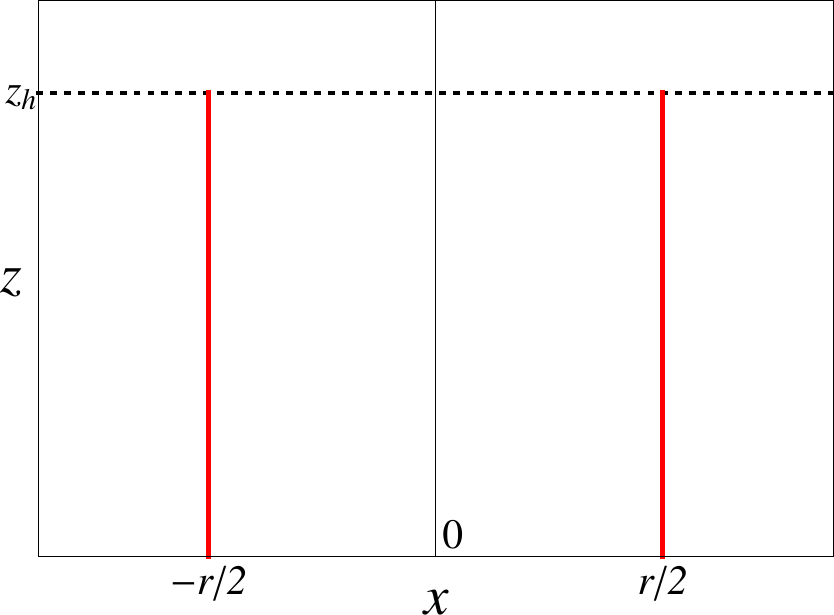}
\label{fig:wsb}
}\\\vspace*{-0.3cm}
\subfigure[confined]{
\includegraphics[scale=0.32]{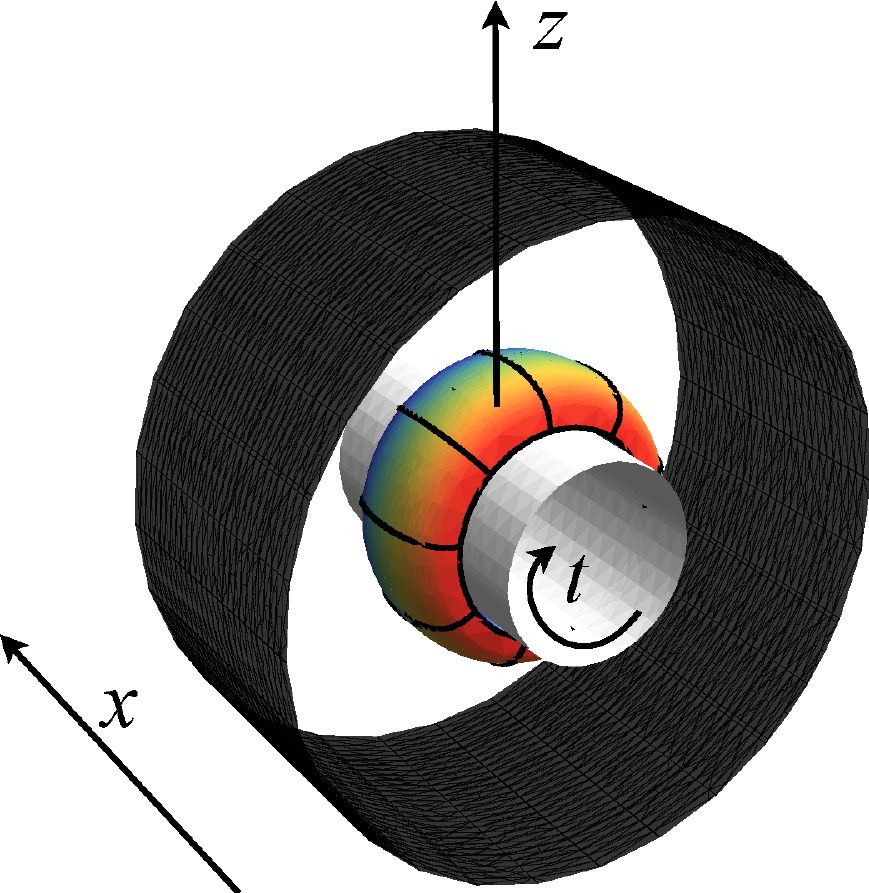}
\label{fig:ws}
}
\hspace*{.2cm}
\subfigure[deconfined]{
\includegraphics[scale=0.20]{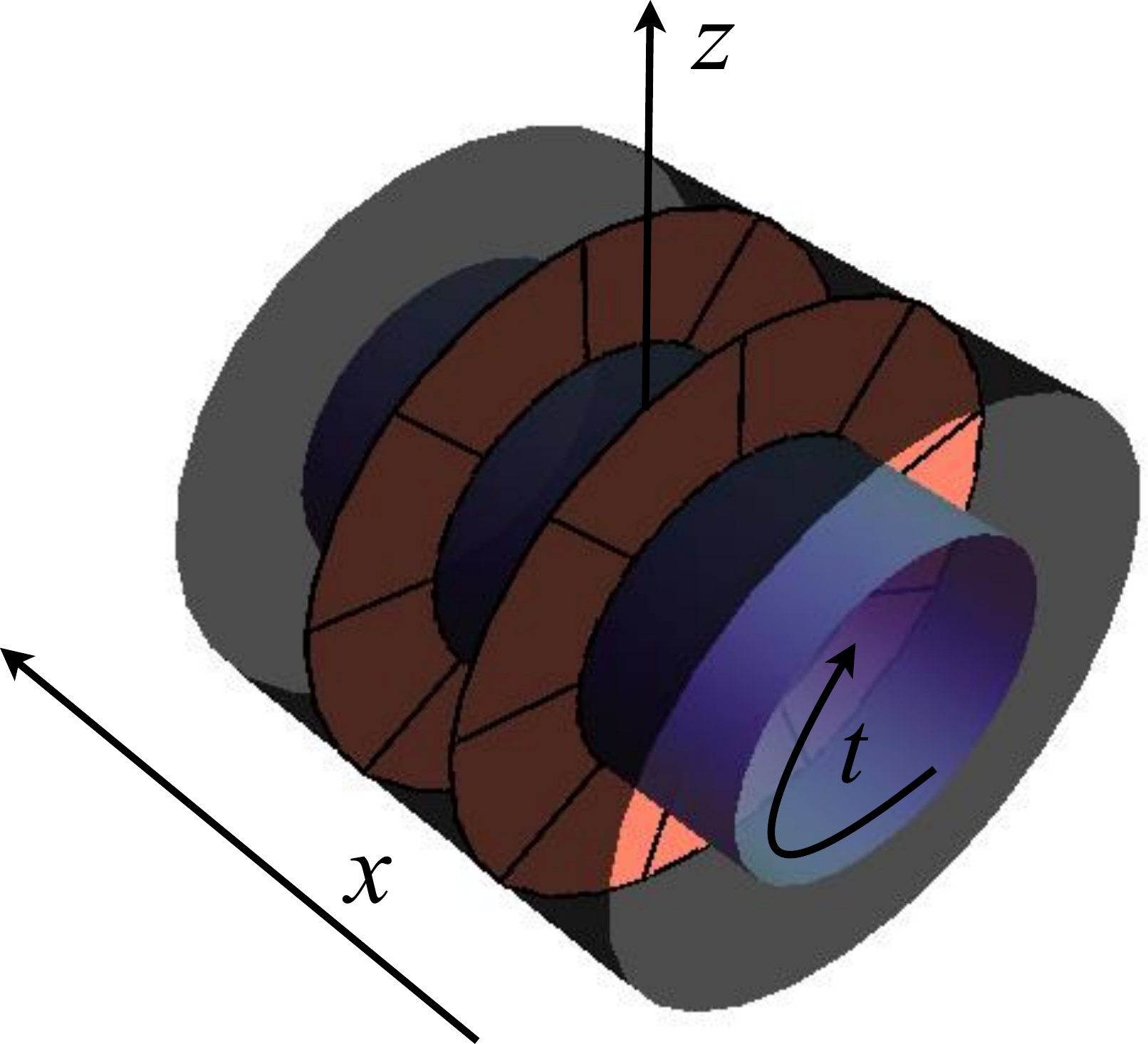}
\label{fig:wss}
}
\caption{String configurations corresponding to the confined  (a,c) and  deconfined phase (b,d). In (a,c) there is one string with endpoints at the positions of the static color sources, $x=\pm r/2$ at  $z=0$, which does not intersect the horizon (dark cylinder in (c)); in (b,d) there are two strings stretched between $z=0$ and the horizon $z=z_h$, with $x(z)=\pm r/2$.}\label{fig:ws}
\end{figure}

Let us now turn to the problem formulated at the beginning of this study.
At finite temperature, the free energy $F(r,T)$ of an infinitely heavy quark-antiquark pair at distance $r$ can be obtained in QCD from the correlation function of two Polyakov loops:
\begin{equation}\label{twopolyakov}
\langle{\CMcal P}(\vec x_1) {\CMcal P}^\dagger(\vec x_2) \rangle= e^{-\frac{1}{T} F(r,T)+\gamma(T)}
\end{equation}
with $r=|\vec x_1-\vec x_2|$ and $\gamma(T)$ a normalization constant. Moreover, the expectation value of a single Polyakov loop
\begin{equation}\label{polyakov}
\langle{\CMcal P} \rangle= e^{- \frac{1}{2 T}  F^\infty(T)}
\end{equation}
($ F^\infty(T)=F(r=\infty,T)$ and neglecting the normalization) is the order parameter for the deconfinement transition in a pure $SU(N)$ theory \cite{polyakov}.
Within the gauge/string duality approach, we can attempt a calculation of the expectation values in \eqref{twopolyakov} and \eqref{polyakov} considering string configurations in the 5D manifold having the Polyakov loops as boundary, looking at  the configurations of minimal surfaces and computing  their worldsheet action. 
The worldsheet action is the Nambu-Goto one
\begin{equation}\label{NG}
S_{\mbox{\tiny NG}}=\frac{1}{2\pi \alpha^\prime} \int d^2\xi \sqrt{\det\left[g_{MN}\left(\partial_a X^M\right)\left(\partial_b X^N\right)\right]}
\end{equation}
for a string with endpoints attached to the positions of the heavy quark and antiquark, $x=\pm r/2$ at $z=0$, so that \cite{Maldacena:1998im,Andreev:2006nw}
\begin{equation}\label{free-en}
F(r,T)=TS_{\mbox{\tiny NG}}. 
\end{equation}
$X^M$'s are the coordinates of the five dimensional spacetime, $g_{MN}$ the metric tensor associated to the line element \eqref{metric}, and the determinant is taken over the indices $a,b=0,1$ labeling the worldsheet coordinates $\xi^a$.
We parametrize them with $\xi^0=t$ and $\xi^1=x$, and look for a static solution $z(x)$ satisfying the conditions  $z(x=0)=z_0$, $z^\prime (x=0)=0$ and $z(x=\pm r/2)=0$, the prime denoting the derivative with respect to $x$.
Examples of such configurations are shown in Fig.~\ref{fig:wsa}-\ref{fig:wss}.

From \eqref{NG} and \eqref{free-en} we obtain an expression for the free energy:
\begin{equation}\label{Ftemp}
F(r,T)=\frac{g}{\pi}\int_{-r/2}^0 dx \frac{e^{c^2z^2}}{z^2}\,\sqrt{f(z)+\left(z^\prime\right)^2}\,\,
\end{equation}
with $g=\frac{R^2}{\alpha^\prime}$. Moreover, an equation of motion follows from \eqref{Ftemp}, with the first integral
\begin{equation}
{\CMcal H}=\frac{e^{c^2z^2}}{z^2}\,\frac{f(z)}{\sqrt{f(z)+\left(z^\prime\right)^2}}
\end{equation}
which allows us to express $F(r,T)$ in terms of $z_0$ and $f_0=f(z_0)$.
Defining $v=z/z_0$, after having subtracted the UV ($v\to0$) divergence (corresponding to subtract the infinite quark and antiquark mass in four dimensional QCD\cite{Maldacena:1998im}), we obtain
\begin{equation}\label{F}
\hat{F}(\lambda)=\frac{g}{\pi\lambda}\left[-1+\int_0^1\frac{dv}{v^2}\left(\frac{e^{\lambda^2 v^2}}{\tau(v)}-1\right)\right]
\end{equation}
where $\hat F=F/c$,  $\lambda=c\,z_0$ and
\begin{eqnarray}
\tau(v) & = & \sqrt{1-\frac{f_0}{f(z_0 v)}\,v^4e^{2\lambda^2\left(1-v^2\right)}}\,\,.
\end{eqnarray}
The distance $\hat  r=c r$ can also be expressed parametrically, 
\begin{equation}\label{r}
\hat r(\lambda)=2 \lambda\sqrt{f_0} \int_0^1 dv\, \frac{v^2\, e^{\lambda^2(1-v^2)}}{\tau(v)f(z_0 v)}\,\,,
\end{equation}
so that, from \eqref{F} and \eqref{r} we can write the free energy as a function of the temperature $\hat T=T/c$, for a range of values of the chemical potential $\hat\mu=\mu/c$.
We fix the parameters $\kappa$ and $g$ to one, postponing to the end the discussion on the numerical results.
%
\begin{figure}
\subfigure[confined]{
\includegraphics[scale=0.45]{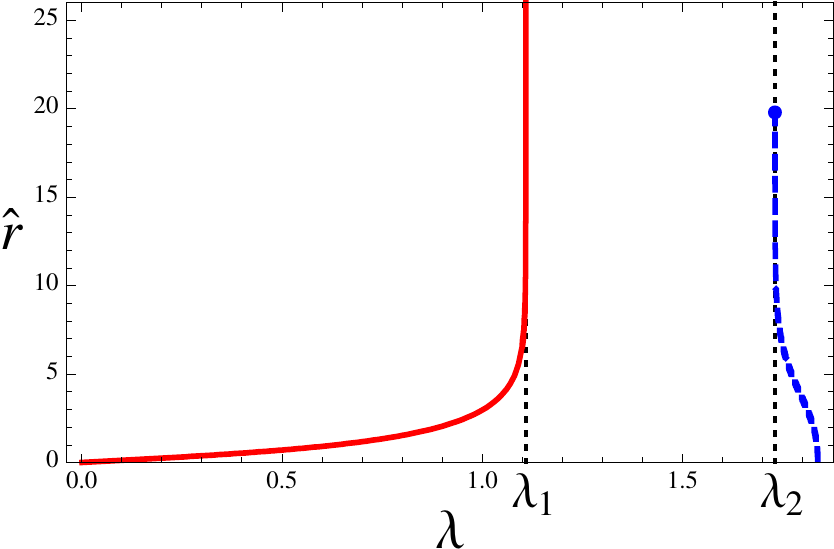}
\label{fig:r1a}
}
\hspace*{.2cm}
\subfigure[deconfined]{
\includegraphics[scale=0.45]{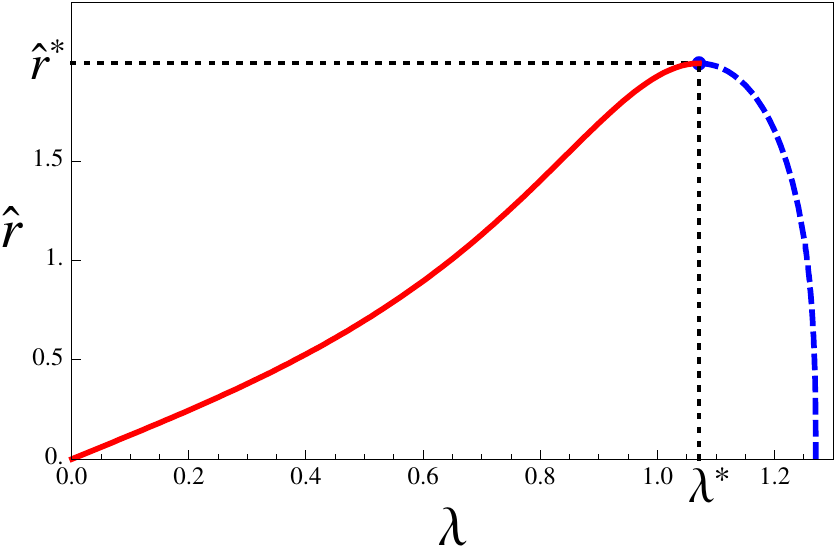}
\label{fig:r1b}
}\\ \vspace*{0.0cm}
\subfigure[confined]{
\includegraphics[scale=0.45]{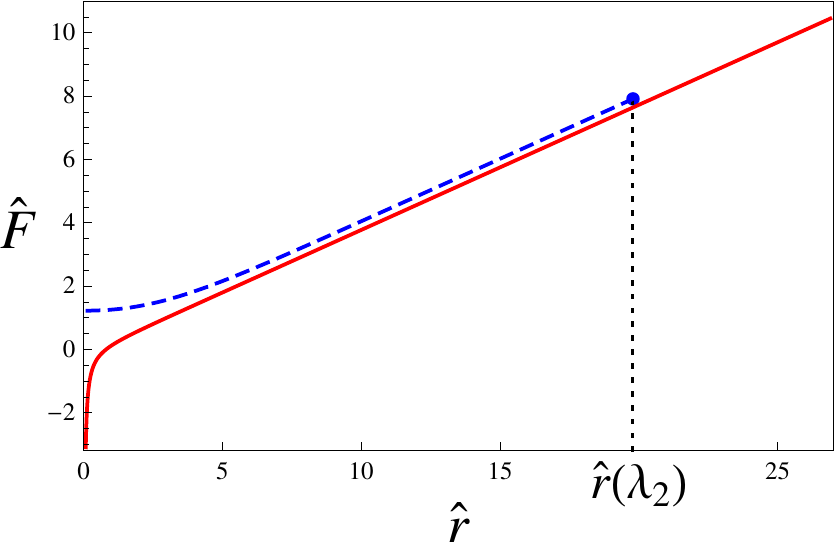}
\label{fig:r1c}
}
\hspace*{.2cm}
\subfigure[deconfined]{
\includegraphics[scale=0.45]{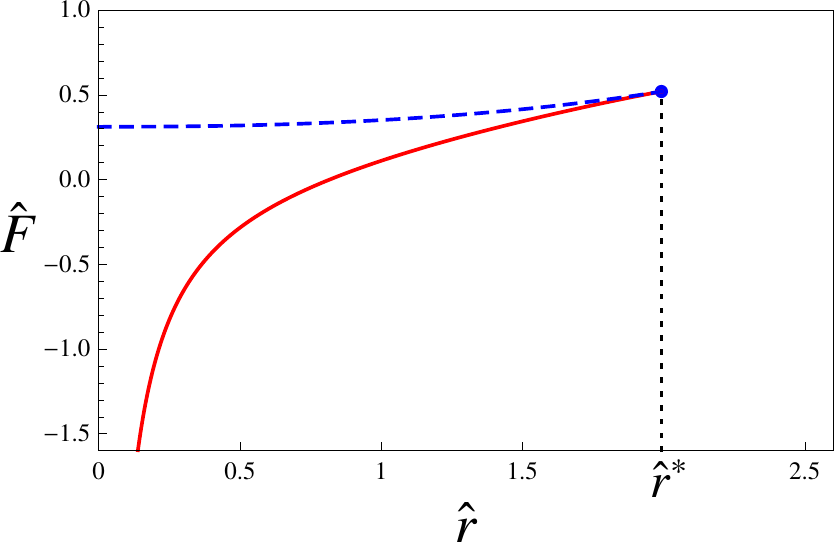}
\label{fig:r1d}
}
\caption{Interquark distance $\hat r(\lambda)$ in the confined (a) and deconfined phase (b). 
In (a) (here for $\hat\mu=0.5$ and $\hat T=0.82T^*$), for $0\leqslant\lambda\leqslant\lambda_1$,  $\hat r(\lambda)$ spans the whole real axis (red solid line); as $\lambda_2\leqslant\lambda\leqslant\hat z_h$, $\hat r(\lambda)$ monotonically decreases from a finite value $\hat r(\lambda_2)$ to $\hat r(\hat z_h)=0$ (blue dashed line). 
The free energies corresponding to the two branches are plotted in (c). 
In (b) (here for $\hat\mu=0.5$ and $\hat T=1.65T^*$) $\hat r(\lambda)$ has a maximum for $\lambda_1=\lambda_2=\lambda^*$; the free energy is plotted in (d).}\label{fig:r1}
\end{figure}

As for $\hat\mu=0$ \cite{Andreev:2006nw}, also for finite chemical potential the distance $\hat r(\lambda)$, written parametrically in \eqref{r}, has two branches, shown in Fig.~\ref{fig:r1a}.
For $0\leqslant\lambda\leqslant\lambda_1$,  $\hat r$ increases monotonically from $\hat r(0)=0$, spans the whole real axis and diverges when $\lambda\to\lambda_1$.
For $\lambda_2\leqslant\lambda\leqslant\hat z_h=c z_h$,  $\hat r$ monotonically decreases from a finite value $\hat r(\lambda_2)$ down to $\hat r(\hat z_h)=0$; for $\lambda_1<\lambda<\lambda_2$,   $\hat r$ has an imaginary part.
Selecting the branch $0\leqslant\lambda\leqslant\lambda_1$, we can compute $\hat F$ for all distances, with the result depicted in Fig.~\ref{fig:r1c}. 
As shown in the same figure, the values of the free energy corresponding to the branch $\lambda_2\leqslant\lambda\leqslant\hat z_h$ are larger than those for $0\leqslant\lambda\leqslant\lambda_1$.
The resulting free energy can be parameterized similarly to the Cornell expression for the static quark-antiquark potential:
\begin{equation}
\hat F=-\frac{a}{\hat r}+\hat b+ {\hat \sigma}^2 \hat r \,\,\, .
\end{equation}
The coefficient $a$ turns out to be essentially independent of $T$ and $\mu$; on the other hand, at fixed $\hat T=0.05$, the parameter $\hat b$ increases, from $\hat b=-0.073$ at $\hat\mu=0$ to $\hat b=-0.051$ at $\hat\mu=0.6$, while $\hat\sigma$ decreases from $\hat\sigma=0.64$ to $\hat\sigma=0.62$ when $\hat\mu$ is changed from $\hat\mu=0$ to $\hat\mu=0.6$.
  
For fixed values of the chemical potential $\hat \mu=\hat\mu^*$, increasing $\hat T$, the two points $\lambda_1$ and $\lambda_2$ along the $\lambda$ axis get closer to each other, and at a certain value $\hat T^*$ they coincide: $\lambda_1=\lambda_2=\lambda^*$. 
For $\hat T\geqslant\hat T^*$,  $\hat r(\lambda)$ is real for $0\leqslant\lambda\leqslant\hat z_h$, and never exceeds the value $\hat r(\lambda^*)$, as shown in Fig.~\ref{fig:r1b}: at the point $(\hat\mu^*,\hat T^*)$ in the $\hat\mu-\hat T$ plane $\hat r(\lambda)$ is always bounded.
Therefore, taking $\hat\mu=\hat\mu^*$ and $\hat T\geqslant\hat T^*$, $\hat F$ can be evaluated only for $\hat r\leqslant\hat r(\lambda^*)$, Fig.~\ref{fig:r1d}.
For $\hat r>\hat r(\lambda^*)$ there is another string configuration, \emph{i.e.} the one with two strings stretched between the boundary $\hat z=0$ and the black-hole horizon $\hat z=\hat z_h$, plotted in Figs. \ref{fig:wsb} and \ref{fig:wss} \cite{Rey:1998bq}.
This configuration holographically represents two deconfined quarks. To evaluate the corresponding action we choose a different parametrization of the worldsheet: $\xi^0=t$ and $\xi^1=z$, obtaining the regularized free energy
\begin{equation}\label{Finf}
\hat F^\infty=
\frac{g}{\pi}\left[-\frac{1}{\hat z_h}+\int_0^{\hat z_h}\frac{d\hat z}{\hat z^2}\left(e^{\hat z^2}-1\right)\right]+\zeta(\hat\mu,\hat T)
\end{equation}
with $\hat z=cz$; $\zeta(\hat\mu,\hat T)$ is a constant related to the regularization procedure, and fixed by matching $\hat F(\hat r^*)=\hat F^\infty$.
In this case the free energy does not depend on $\hat r$, but only on the temperature and chemical potential.
\begin{figure}
\includegraphics[scale=0.68]{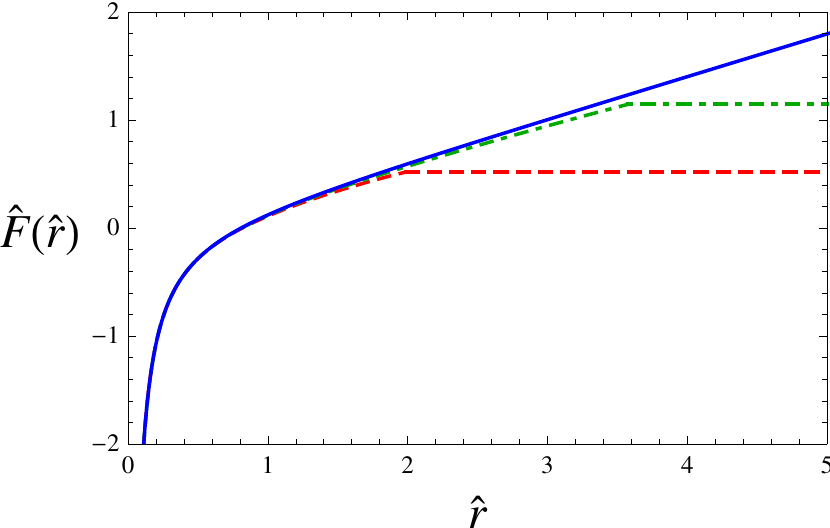}
\caption{Free energy in the confined phase, $\hat\mu=0.5$, $\hat T=0.82T^*$ (blue solid line), and in the deconfined one, for $\hat\mu=0.5$, $\hat T=1.23T^*$ (green dot-dashed) and $\hat\mu=0.5$, $\hat T=1.65T^*$ (red dashed). For $\hat\mu=0.5$ the critical temperature is $\hat T^*=0.122$. }\label{fig:F}
\end{figure}

As a result, the free energy is depicted in Fig.~\ref{fig:F}. 
At large values of $\hat r$, for a given value of the chemical potential $\hat\mu$ there is a temperature $\hat T^*(\hat\mu)$ such that for $\hat T$ less than $\hat T^*$ the free energy linearly grows with $\hat r$, while for $\hat T\geqslant\hat T^*$ the solution coming from the connected string configuration (Fig.~\ref{fig:wsa}) is only valid up to $\hat r=\hat r(\lambda^*)$; from that point on, $\hat F$ is constant \eqref{Finf}.
This behavior of the free energy with distance and temperature has been observed in quenched lattice simulations at $\hat\mu=0$ \cite{Kaczmarek:2005ui}.

\begin{figure}
\includegraphics[scale=0.7]{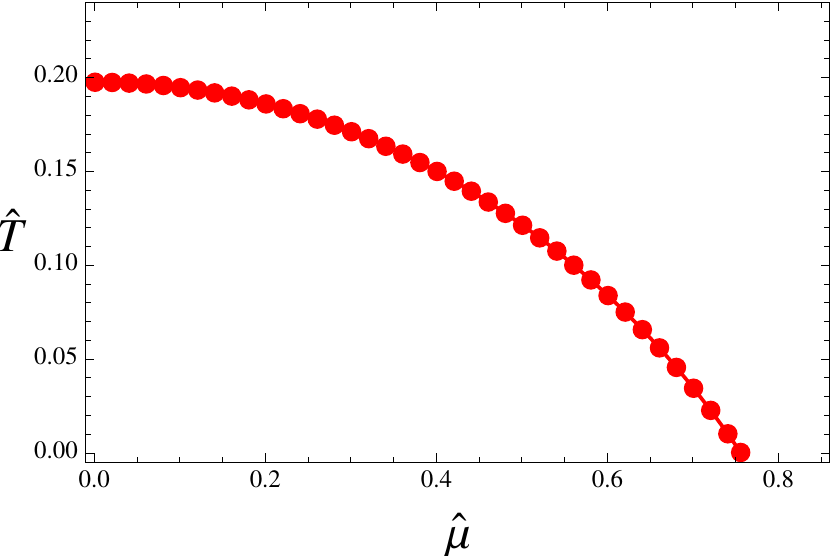}
\caption{Deconfinement transition line in the $\hat\mu-\hat T$ plane. 
The line divides the plane in two regions, a hadron phase near the origin, and a deconfined phase beyond the curve.}\label{fig:phasediag}
\end{figure}

In the $\hat\mu-\hat T$ plane, the curve defined by the points $(\hat\mu,\hat T^*(\hat\mu))$ is plotted  in Fig.~\ref{fig:phasediag}.  
This can be considered as a picture of the deconfinement transition in the QCD  phase diagram. 
For low values of $T$ and $\mu$ there is a confined phase characterized by a linearly increasing interquark potential.
Outside this region, a finite amount of energy is sufficient to separate the quark-antiquark pair, as expected in a deconfined phase.
The  picture agrees with the diagram obtained by, \emph{e.g.}, Nambu-Jona-Lasinio \cite{Buballa:2003qv} and other effective models \cite{Schaefer:2004en}.

\begin{figure}[hb]
\includegraphics[scale=0.72]{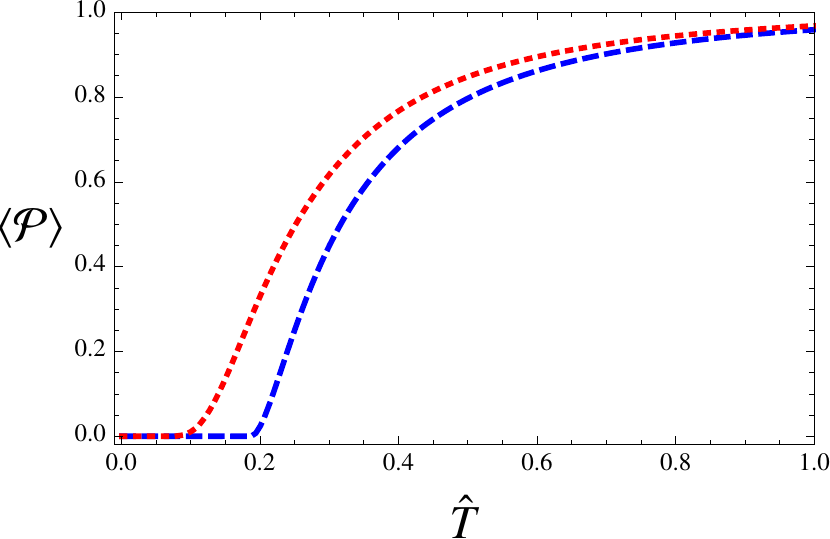}
\caption{Polyakov loop $\langle{\CMcal P}\rangle$ versus $\hat T$ (normalized at its value for $T\to\infty$) for $\hat\mu=0.2$ (blue, dashed, right curve) and $\hat\mu=0.7$ (red, dotted, left curve).}\label{poloop}
\end{figure}

In the same holographic framework, the Polyakov loop $\langle{\CMcal P}\rangle$ can be computed by Eq.\eqref{polyakov}.
In the confined phase, $\hat F(\hat r)$ diverges as $\hat r\to\infty$, so $\langle{\CMcal P}\rangle$ vanishes, while in the deconfined one it is determined by $\hat F^\infty$ from the configuration in Figs.~\ref{fig:wsb} and \ref{fig:wss}.
The result is depicted in Fig.~\ref{poloop}: for each value of $\hat\mu$, the Polyakov loop vanishes as $\hat T<\hat T^*$; it starts growing at $\hat T=\hat T^*$ \cite{andreev-pol}.
The smooth increase can be interpreted as a continuous transition between the hadron and the deconfined phase.
%
%


As pointed out in \cite{Lee:2009bya}, the parameter $\kappa$ scales as $\kappa\sim\sqrt{N_c}$ and its numerical value is model dependent. 
If we take the limit $N_c\to\infty$, the deconfinement line in Fig.~\ref{fig:phasediag} becomes flat, with $\hat T^*$ not depending on $\hat\mu$, in agreement with the result found in \cite{McLerran:2007qj}.
Considering a finite $\kappa$ in \eqref{rel}, the deconfinement line has the shape shown in Fig.~\ref{fig:phasediag}.

In order to discuss numerical results, we need to choose the values of the parameter $\kappa$ in (\ref{rel}) and of the mass scale $c$ appearing in the warp factor.
 We fix  $\kappa\approx 1/2$ from the ratio $\hat T_c/\hat\mu_c\approx 0.5$ found in \cite{Buballa:2003qv}, with $\hat T_c=\hat T^*$ at $\hat\mu=0$ and $\hat\mu_c=\hat\mu^*$ at $\hat T=0$
%
%
($\kappa$ could also be determined in this soft wall model studying different observables,  as done in \cite{Lee:2009bya} for the hard wall model).
The mass scale $c$ induced by the 
warp factor
can be fixed from the result $T_c\approx 170 $ MeV \cite{referenza}:  $c\approx 850$ MeV. 
On the other hand, the value of this scale from the $\rho$ meson mass is $c\approx  670$ MeV \cite{Andreev:2006vy}, which gives $T_c\approx 134$ MeV and $\mu_c\approx 248$ MeV.  
\footnote{Values of the scale $c$ have also been obtained in the framework of light front holography, $c\simeq 540$ MeV from the $\rho$ meson spectrum, and $c \simeq 500$ MeV
from the baryon spectrum  \cite{deTeramond:2009xk}. The corresponding
critical temperature at $\mu=0$ turns out to be quite low.}

The numerical values of the critical chemical potential are a remarkable result of our study.
The other result is that the gauge/string duality approach can be used to obtain information on the 
deconfinement transition in QCD, finding an agreement with the outcome of effective theories (even if the model is not sensitive to the structure of the vacuum). 
The success of the holographic framework  is  noticeable, considering the difficulty of lattice QCD to explore the phase diagram along the axis of  the chemical potential, and opens interesting perspectives for the analysis, for example,  of the superconducting phases of QCD. 

\vspace*{1cm}
\begin{acknowledgments}
We are grateful to Fulvia~De~Fazio and Marco~Ruggieri for  discussions. 
This work was supported, in part, by the EU contract No. MRTN-CT-2006-035482, ``FLAVIAnet". 
FG was supported, in part, by the grant ``\emph{Borse di ricerca in collaborazione internazionale}'' by \emph{Regione Puglia, Italy}; she thanks the IPPP, Durham, for hospitality during the completion of this work.
\end{acknowledgments}

\end{document}